\documentclass[aps, prb, reprint, superscriptaddress, floatfix]{revtex4-2}
\usepackage{graphicx}
\usepackage{dcolumn}
\usepackage{bm}
\usepackage{physics}
\usepackage{amsfonts}
\usepackage{mathrsfs}
\usepackage{subfigure}
\usepackage[svgnames]{xcolor}
\usepackage[colorlinks=true,linkcolor=NavyBlue,anchorcolor=red,citecolor=NavyBlue, urlcolor=NavyBlue]{hyperref}
\usepackage{color}
\usepackage{xcolor}
\DeclareMathAlphabet{\pazocal}{OMS}{zplm}{m}{n}
\usepackage{amsmath}
\usepackage{amssymb}
\usepackage[T1]{fontenc}
\begin{document}

\preprint{APS/123-QED}
 
\title{Suppressed coarsening after an interaction quench in the Holstein chain}

\author {Ho Jang}
\affiliation{Department of Physics, University of Virginia, Charlottesville, Virginia, 22904, USA}

\author {Gia-Wei Chern}
\affiliation{Department of Physics, University of Virginia, Charlottesville, Virginia, 22904, USA}

\begin{abstract}
We investigate the nonequilibrium dynamics induced by an interaction quench in the semiclassical Holstein model within the Ehrenfest nonadiabatic framework, which describes an isolated hybrid quantum–classical system with strictly conserved total energy. Focusing on the half-filled case, where the equilibrium ground state exhibits commensurate charge-density-wave (CDW) order for any nonzero coupling, we identify three distinct post-quench dynamical regimes as a function of the final electron-phonon coupling: a nonequilibrium metallic state without CDW order, an intermediate regime characterized by slow scale-invariant ordering dynamics, and a frozen CDW state with arrested coarsening and immobile kinks. We analyze the intermediate regime in detail and uncover an unconventional algebraic decay of the kink density, $n\sim t^{-1/3}$, distinct from both ballistic annihilation and diffusive coarsening in classical dissipative systems. We show that this anomalous exponent arises from the hybrid nature of the dynamics: while the lattice evolves deterministically, the electronic degrees of freedom act as an effective internal bath that induces diffusive kink motion without energy dissipation. An effective reaction-diffusion description, incorporating both annihilation and elastic scattering of kinks, quantitatively accounts for the observed scaling behavior. Our results reveal a distinct coarsening mechanism in isolated hybrid systems, demonstrating how internal quantum dynamics can qualitatively reshape defect kinetics far from equilibrium.\end{abstract}

\date{\today}
\maketitle

\date{\today}
\maketitle

\section{Introduction}
\label{sec:intro}

The nonequilibrium relaxation of symmetry-broken phases following a rapid quench has long been a central problem in statistical physics~\cite{Bray1994,Cugliandolo15,Onuki2002,Puri2009,hohenberg1977,Tauber2014}. In the classical setting, this process is described by phase-ordering kinetics, in which ordering proceeds through the formation and growth of domains corresponding to competing equilibrium phases, separated by topological defects such as domain walls or vortices. At late times, the dynamics is governed by the slow motion and annihilation of these defects and becomes largely insensitive to microscopic details, giving rise to dynamical scaling characterized by a single growing length scale $L(t)$. The domain size typically follows a power-law growth, $L\sim t^{\alpha}$, with the exponent $\alpha$ determined by conservation laws and the nature of the order parameter. Canonical examples include curvature-driven Allen-Cahn dynamics for nonconserved order parameters~\cite{Allen1972}, yielding $\alpha=1/2$, and diffusion-limited Lifshitz-Slyozov-Wagner coarsening for conserved order parameters~\cite{Lifshitz61,Lifshitz62,Wagner61}, with $\alpha=1/3$. In thermal quenches, dissipation to an external bath provides the arrow of time and ultimately drives the system toward equilibrium.

A complementary direction that has attracted increasing attention concerns the nonequilibrium evolution of isolated many-body systems, both quantum and classical~\cite{polkovnikov2011,mitra2018,marino2022,Cugliandolo_2017,Cugliandolo2018}. This interest has been driven in large part by ultracold-atom experiments, where coupling to the environment can be made negligibly weak~\cite{bloch2005}. In this context, interaction quenches serve as paradigmatic realizations of unitary or Hamiltonian dynamics, in which the total energy is strictly conserved. This immediately raises a fundamental question: to what extent can concepts such as coarsening, defect-controlled growth, and dynamical universality---originally developed for dissipative systems---be meaningfully extended to isolated settings?

A substantial body of experimental and theoretical work has shown that isolated many-body systems can display extended nonequilibrium regimes with scaling, aging, and a growing correlation length following interaction quenches~\cite{greiner2002,lucia2010,trotzky2012,langen2013,meinert2013,cheneau2012,gring2012,ronzheimer2013,nicklas2015,prufer2018,manovitz2025,damle1996,mukerjee2007,lamacraft2007,williamson2016,williamson2016b}. In contrast to dissipative systems, conservation laws---most notably energy conservation---strongly constrain relaxation and can preclude equilibration to a thermal Gibbs ensemble~\cite{Rigol2007,Berges2008,Kollar2011,bertini2015,Chiocchetta2016,bertini2016}. Instead, the dynamics may approach nonthermal or prethermal stationary states, or exhibit slow, coarsening-like evolution toward critical or quasi-ordered configurations.
Analytically tractable models and controlled approximations further demonstrate that such isolated coarsening can occur without thermalization, giving rise to slow, scale-invariant dynamics near dynamical critical points~\cite{chandran2013,sciolla2013,gagel2015,berges2015,maraga2015,goo2022}. The emergence of aging and universal scaling depends sensitively on dimensionality and conservation laws, and in some cases is restricted to dimensions that support finite-temperature order in equilibrium. From this perspective, coarsening in isolated systems reflects the appearance of nonthermal fixed points and nonequilibrium universality classes distinct from conventional thermal relaxation.

One-dimensional systems occupy a special place in this emerging picture, as both equilibrium and nonequilibrium ordering are profoundly influenced by fluctuations and topology~\cite{Vachaspati2006}. Following a classical thermal quench, coarsening dynamics in one dimension is governed by the motion and annihilation of topological defects, or kinks, separating locally ordered domains, as exemplified by Ising and Potts chains. For nonconserved order-parameter dynamics (e.g., Glauber dynamics in the Ising model), kinks undergo unbiased diffusion and annihilate upon encounter, leading to diffusion-limited growth of the characteristic domain size $L\sim t^{1/2}$~\cite{Toussaint83,Torney83,Krapivsky_book,Ovchinnikov1989,Hughes1995RandomWalks}. In contrast, when the order parameter is conserved, as in Kawasaki dynamics, coarsening proceeds via the diffusion of entire domains through particle exchange, resulting in a slower growth law $L\sim t^{1/3}$~\cite{kawasaki1982, kawasaki1983}. These scaling behaviors underscore the dominant role of defect kinetics and conservation laws in one dimension, where strong fluctuations preclude true long-range order at finite temperature and place coarsening firmly under kink-controlled dynamics.

Recent theoretical work has shown that coarsening in isolated or Hamiltonian one-dimensional systems can differ qualitatively from its classical dissipative counterpart, with energy conservation fundamentally reshaping late-time ordering dynamics~\cite{dziarmaga2005,moca2017,kormos2016,bertini2019,alba2017,james2019,kou2023}. Rather than being governed by diffusive relaxation, isolated one-dimensional systems can exhibit long-lived nonequilibrium regimes in which ordering is controlled by the coherent propagation and interaction of kinks. In integrable or near-integrable settings, such as quantum Ising chains, stable quasiparticles or extensive conservation laws can strongly suppress kink annihilation, arresting coarsening or leading to nonthermal steady states with finite domain-wall densities.

Beyond integrable or confined limits, coarsening in isolated one-dimensional systems is governed by kink kinetics that differ from classical diffusion-controlled dynamics~\cite{Staniscia2019,bastianello2020}. In Hamiltonian quenches of classical $\phi^4$ and Ising-type models, kinks propagate ballistically and undergo frequent elastic scattering, while annihilation events remain rare. At long times, annihilation nevertheless dominates, yielding an algebraic decay of the kink density $n\sim 1/t$ and a linear growth of the characteristic length scale $L\sim t$. Energy conservation further enriches the dynamics, producing scaling regimes absent in dissipative systems through a coupling between kink motion and a slowly evolving local temperature field.

In this work, we study the nonequilibrium dynamics following an interaction quench in a hybrid quantum--classical system, namely the semiclassical Holstein model~\cite{Holstein59, holstein1959b}. Within the Ehrenfest nonadiabatic framework, the post-quench evolution is strictly Hamiltonian and conserves the total energy: the lattice degrees of freedom evolve under classical equations of motion, while the quadratic electronic subsystem is governed by the von~Neumann equation for the single-particle density matrix. We focus on the half-filled case, which supports a commensurate charge-density-wave (CDW) ground state with a doubled unit cell for any nonzero electron--lattice coupling~\cite{Gruner1988, Gruner1994}. Depending on the quench parameters, we identify three distinct dynamical regimes: (i) a nonequilibrium metallic state without CDW order, (ii) an intermediate regime displaying slow, scale-invariant ordering dynamics, and (iii) a frozen CDW state characterized by immobile kinks and arrested dynamics.

We concentrate on the intermediate quasi-coarsening regime and analyze its domain-growth kinetics in detail. We find an unconventional algebraic decay of the kink density, $n\sim t^{-1/3}$, which is markedly different from both ballistic annihilation and purely diffusive coarsening in classical dissipative systems. This anomalous exponent originates from the hybrid nature of the dynamics: although the lattice evolves deterministically, the electronic degrees of freedom act as an effective internal bath that induces diffusive motion of kinks. We show that the observed $1/3$ power law can be captured by an effective kink kinetics model in which kinks perform unbiased random walks and annihilate upon collision with high probability, while elastic scattering occurs with a finite probability and slows the coarsening process. These results reveal a distinct coarsening mechanism in isolated hybrid systems, where internal quantum degrees of freedom qualitatively reshape late-time defect dynamics.

The remainder of this paper is organized as follows. In Sec. II, we introduce the semiclassical Holstein model and describe its formulation within the Ehrenfest nonadiabatic dynamics framework, which provides a canonical setting for hybrid quantum–classical systems with conserved total energy. In Sec. III, we present the dynamical phase diagram obtained from our nonadiabatic simulations and identify three distinct post-quench regimes: a nonequilibrium metallic state, an intermediate quasi-coarsening regime, and a frozen regime characterized by arrested coarsening and immobile kinks. Section IV is devoted to an empirical reaction–diffusion model for kink dynamics that accounts for the observed power-law behavior in the quasi-coarsening phase. Finally, Sec. V summarizes our results and discusses their implications and possible directions for future work.

\section{Nonadiabatic Dynamics in Hybrid Quantum-Classical Systems}

\label{sec:holstein}

In this section, we first formulate the semiclassical Holstein model within an energy-conserving Ehrenfest nonadiabatic framework~\cite{marx2009,Li2005}, and motivate its use as a platform for studying nonequilibrium coarsening dynamics beyond both purely classical and fully quantum limits. In classical Hamiltonian field theories, such as the one-dimensional $\phi^4$ model, recent work has shown that energy conservation can lead to reaction-dominated kink annihilation and unconventional coarsening laws. Extending these ideas to genuinely quantum systems, however, remains challenging: exact treatments are largely restricted to integrable limits, while simulations of generic nonintegrable models are severely constrained by accessible system sizes and timescales. As a result, large-scale coarsening phenomena in quantum many-body systems are difficult to address directly.

The semiclassical Holstein model circumvents these limitations by combining quantum electronic dynamics with classical lattice degrees of freedom, while preserving strict energy conservation under nonequilibrium driving. Within the Ehrenfest approach, the electronic subsystem evolves according to the von Neumann equation, while the lattice follows Hamiltonian equations of motion, allowing for numerically exact simulations of very large systems over long times. This hybrid quantum–classical structure provides access to coarsening regimes far beyond the reach of fully quantum approaches, while remaining physically realistic, as the lattice variables directly represent atomic distortions in electron–phonon coupled materials. In this sense, the semiclassical Holstein model offers a controlled and experimentally relevant setting for exploring coarsening dynamics in nontrivial interacting many-body systems.

We consider the spinless Holstein model at half filling on a one-dimensional lattice~\cite{Holstein59, holstein1959b}, described by the Hamiltonian
\begin{eqnarray}
	& & \hat{\mathcal{H}} = -t_{\rm nn}\sum_{\langle ij\rangle} \left(\hat{c}^{\dagger}_{i}\hat{c}_{j} + \hat{c}^{\dagger}_{j}\hat{c}_{i}\right) 
	-g \sum_i \left(\hat{n}_i - \frac{1}{2}\right)\hat{Q}_i \nonumber \\
	& & \qquad + \sum_i \left( \frac{\hat{P}_i^2}{2m} + \frac{1}{2}K \hat{Q}_i^2 \right).
\end{eqnarray}
The first term describes nearest-neighbor hopping of spinless fermions with amplitude $t{\rm nn}$, where $\hat{c}_i^\dagger$ ($\hat{c}_i$) creates (annihilates) an electron on site $i$. The last term represents local lattice degrees of freedom, modeled as independent harmonic oscillators, with $\hat{Q}_i$ and $\hat{P}_i$ denoting the lattice displacement and its conjugate momentum at site $i$, respectively. Here $m$ is the phonon mass and $K = m\Omega^2$ defines the bare phonon frequency $\Omega$. The second term describes the local electron-phonon coupling, where $\hat{n}_i = \hat{c}_i^\dagger \hat{c}_i$ is the fermion density operator and $g$ controls the coupling strength. Writing the interaction in a particle-hole symmetric form ensures that, at half filling, the uniform electronic configuration $\langle \hat{n}_i \rangle = 1/2$ corresponds to a vanishing average lattice displacement $\langle \hat{Q}_i \rangle = 0$. This choice eliminates the need for an explicit chemical-potential shift and makes the CDW instability at wavevector $q=\pi$ explicit in the interacting ground state.

The Holstein model is a paradigmatic and extensively studied framework for investigating electron--phonon coupling, capturing the essential physics of itinerant electrons interacting with local lattice distortions. In particular, it has been widely employed to study CDW ordering~\cite{Scalettar1989,noack91,zhang19,chen19,hohenadler19,esterlis19,Weber22,Petrovic24,sankha2024,Yang2025}. Despite its theoretical simplicity, the model also admits a direct physical interpretation as a minimal description of materials with strong local electron--lattice interactions. In this context, the lattice displacement \(Q_i\) represents the amplitude of a local breathing-mode distortion of an \(\mathrm{MO}_6\) octahedron surrounding lattice site \(i\), a structural motif commonly encountered in transition-metal oxides.

On bipartite lattices, including the one-dimensional chain, the half-filled Holstein model exhibits a ground state with long-range charge-density-wave (CDW) order characterized by a doubling of the unit cell. This broken-symmetry state manifests as an imbalance in the electronic density between the two sublattices, $n_{A/B} = \frac{1}{2} \pm \delta n$. Remarkably, this CDW order persists even in the semiclassical limit in which the phonon degrees of freedom are treated classically, demonstrating its robustness beyond purely quantum lattice fluctuations~\cite{esterlis19,Weber22,Petrovic24,sankha2024,Yang2025}. The electronic charge modulation is accompanied by a concomitant static lattice distortion, with opposite displacements on the two sublattices, $Q_{A/B} = \pm \mathcal{Q}$. In one dimension, this commensurate CDW order may be equivalently expressed as a spatially modulated lattice configuration, $Q_i = \mathcal{Q} \cos(q^{\,}_{\rm CDW}\, x_i)$, where the ordering wave vector is fixed by the lattice geometry to $q^{\,}_{\rm CDW} = \pi/a$. In this work, we focus on the nonequilibrium dynamics of the semiclassical Holstein model following a sudden interaction quench into a parameter regime where the equilibrium ground state supports CDW order.

To describe the nonequilibrium dynamics following an interaction quench, we employ the Ehrenfest nonadiabatic framework, which provides a self-consistent hybrid quantum–classical description of the coupled electron–lattice system. Within this approach, the full time-dependent many-body state is assumed to remain factorized into electronic and lattice components, $|\Gamma(t)\rangle = |\Psi(t)\rangle \otimes |\Phi(t)\rangle$. Here $|\Psi(t)\rangle$ denotes the electronic wave function, while $|\Phi(t)\rangle$ describes the phonon sector. The semiclassical approximation is implemented by further assuming that the phonon state factorizes over lattice sites, $|\Phi(t)\rangle = \prod_i |\phi_i(t)\rangle$, so that each lattice site is represented by an independent local phonon wave packet. Within this ansatz, the lattice displacement and momentum are treated as classical dynamical variables defined by instantaneous expectation values $Q_i(t)=\langle \phi_i(t)|\hat{Q}_i|\phi_i(t)\rangle$ and $P_i(t)=\langle \phi_i(t)|\hat{P}_i|\phi_i(t)\rangle$.

The equations of motion for the lattice variables follow from the Heisenberg equations evaluated in the factorized state $|\Gamma(t)\rangle$. Taking expectation values yields a set of coupled equations,
\begin{eqnarray}
\label{eq:newton_eq_app}
    \frac{dQ_i}{dt} &=& \frac{P_i}{m}, \nonumber \\
    \frac{dP_i}{dt} &=& g\, \left(n_i - \frac{1}{2}\right) - K Q_i ,
\end{eqnarray}
which are equivalent to Hamilton’s equations for the classical lattice coordinates and momenta. Here $n_i(t)=\langle \Gamma(t)|\hat{n}_i|\Gamma(t)\rangle=\langle \Psi(t)|\hat{n}_i|\Psi(t)\rangle$, where the reduction to a purely electronic expectation value follows directly from the product structure of the total state. As a result, the lattice motion is driven self-consistently by the instantaneous electronic density profile through an effective, time-dependent potential generated by the electronic subsystem. The lattice dynamics is therefore strictly Hamiltonian and energy conserving, with the electronic degrees of freedom providing a dynamically evolving force that couples the classical lattice motion to the quantum dynamics.

The electronic subsystem evolves according to the time-dependent Schrödinger equation,
\begin{equation}
	i\hbar \frac{\partial}{\partial t} \ket{\Psi(t)} = \hat{\mathcal{H}}_e[\{Q_i(t)\}] \ket{\Psi(t)} ,
\end{equation}
in which the lattice displacements enter as explicitly time-dependent parameters. For the Holstein model, the electronic Hamiltonian is quadratic in the fermionic operators and can be written in terms of a single-particle Hamiltonian matrix, $\hat{\mathcal{H}}_e=\sum_{ij}\hat{c}_i^\dagger H_{ij}\hat{c}^{\,}_j$, with
\begin{eqnarray}
	H_{ij}\bigl[{Q_i(t)} \bigr] = -t_{\langle ij \rangle} - g\, \delta_{ij} Q_i(t)
\end{eqnarray}
Owing to this quadratic structure, an initially Slater-determinant electronic state remains a Slater determinant throughout the time evolution. It is therefore convenient to characterize the electronic degrees of freedom by the single-particle density matrix,
\begin{equation}
\label{eq:rho_def_app}
    \rho_{ij}(t) = \bra{\Psi(t)} \hat{c}^\dagger_j \hat{c}_i \ket{\Psi(t)} ,
\end{equation}
whose diagonal elements yield the local electron densities entering the lattice equations of motion. The time evolution of the density matrix is governed by the von Neumann equation,
\begin{equation}
\label{eq:von_neumann_app}
    i\hbar \frac{d\rho}{dt} = [\rho, H],
\end{equation}
which can be written explicitly as
\begin{eqnarray}
\label{eq:von_neumann_explicit_app}
	i\hbar \frac{d\rho_{ij}}{dt} = \sum_k \left( \rho_{ik} t_{kj} - t_{ik} \rho_{kj} \right) + g \left( Q_j - Q_i \right) \rho_{ij}.
\end{eqnarray}
Together, Eqs.~(\ref{eq:newton_eq_app}) and~(\ref{eq:von_neumann_explicit_app}) define a closed, nonlinear set of equations that couple classical lattice dynamics to quantum electronic evolution on equal footing. These equations constitute the Ehrenfest description of the semiclassical Holstein model and form the basis for all nonequilibrium simulations presented in this work.

From a broader perspective, the Ehrenfest formulation can be viewed as the deterministic, single-trajectory limit of the truncated Wigner approximation~\cite{Paprotzki_2024,tenBrink_2022}, in which quantum fluctuations of the phonon degrees of freedom are neglected. Extensions beyond this limit incorporate such fluctuations by sampling an ensemble of initial phonon configurations, while evolving each realization with the same equations of motion. In this sense, the Ehrenfest approach provides a transparent and controlled baseline for isolating nonadiabatic electron-lattice feedback and the intrinsic nonlinear dynamics of the semiclassical Holstein model.

It is worth emphasizing that, although the Ehrenfest nonadiabatic dynamics is neither purely unitary nor fully classical Hamiltonian dynamics, it nevertheless strictly conserves a total energy,
\begin{eqnarray}
	E = \langle \hat{\mathcal{H}} \rangle = {\rm Tr}[ \rho H] + \sum_i \left( \frac{\hat{P}_i^2}{2m} + \frac{1}{2}K \hat{Q}_i^2 \right).
\end{eqnarray}
Using the coupled equations of motion~(\ref{eq:newton_eq_app}) and~(\ref{eq:von_neumann_app}), one can straightforwardly verify that $dE/dt=0$. This exact energy conservation reflects the Hamiltonian character of the lattice dynamics and the unitary evolution of the electronic subsystem within the Ehrenfest scheme. As a result, the nonequilibrium evolution describes an isolated hybrid quantum-classical system, in which energy is redistributed between electronic and lattice degrees of freedom but never dissipated. This fundamental constraint distinguishes the resulting coarsening dynamics from classical thermal quenches with bath-induced dissipation, placing it in the class of Hamiltonian or microcanonical phase-ordering dynamics.

To facilitate comparison across different interaction strengths, we work with dimensionless parameters defined by natural electronic and lattice scales. A characteristic lattice distortion scale is set by $Q^* \sim g/K$, obtained by balancing elastic and electron-phonon energies. This leads to the dimensionless coupling $\lambda=gQ^*/W=g^2/(WK)$, where $W=4t_{\rm nn}$ is the bandwidth of the one-dimensional tight-binding band.
The nonequilibrium dynamics are further controlled by the intrinsic electronic and lattice timescales. Electronic motion is governed by the hopping amplitude $t_{\rm nn}$, defining the electronic timescale $\tau_{\rm e}=\hbar/t_{\rm nn}$, while lattice dynamics are set by the phonon frequency $\Omega$, with $\tau_{\rm L}=1/\Omega$. Their ratio defines the adiabaticity parameter $r=\tau_{\rm e}/\tau_{\rm L}=\hbar\Omega/t_{\rm nn}$, which interpolates between adiabatic ($r\ll1$) and nonadiabatic regimes. Throughout this work, we fix $r=0.3$ and vary $\lambda$ to isolate the role of electron–phonon coupling in the dynamics. Details of the dimension analysis can be found in Appendix. 

\section{Dynamical Phase diagram}

\begin{figure*}[t!]
    \centering
    \includegraphics[width=0.99\linewidth]{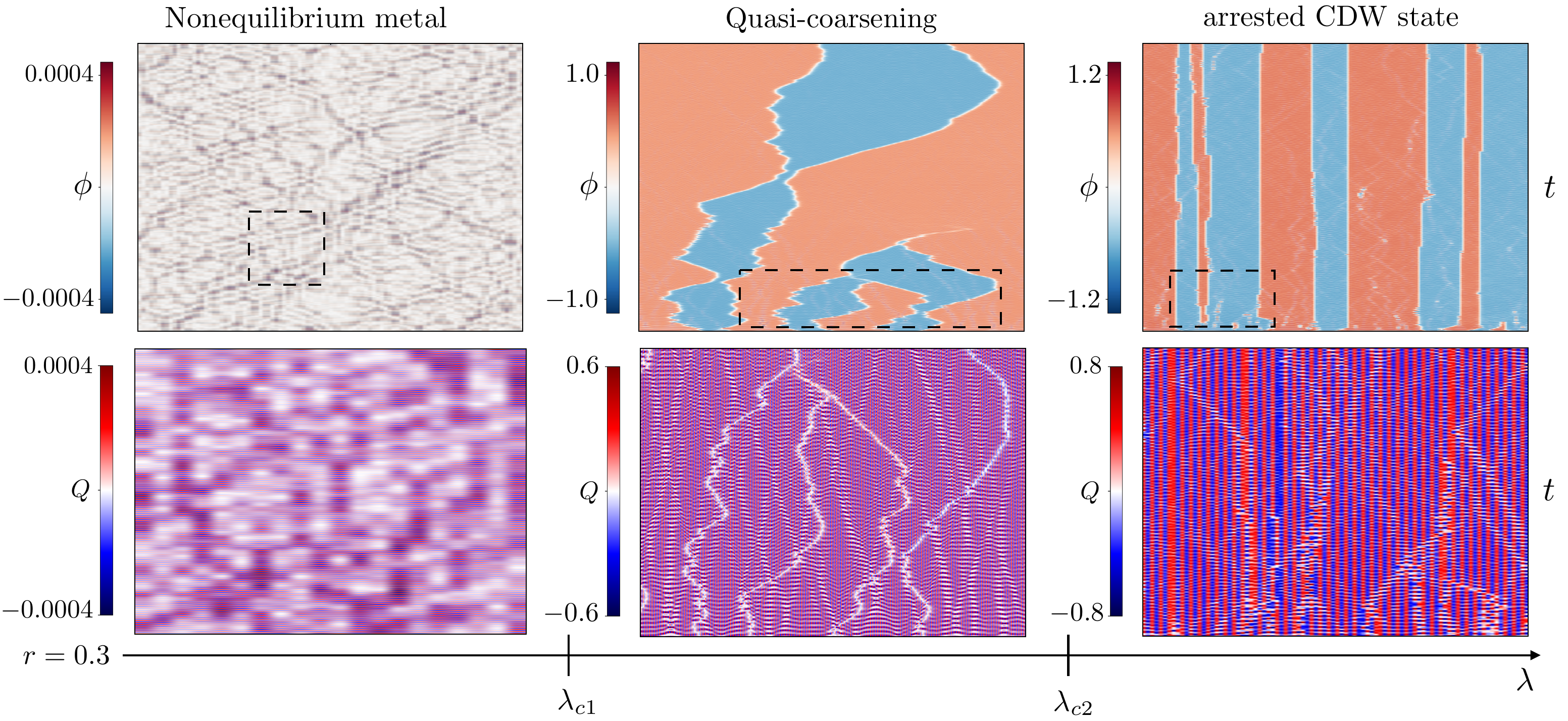}
    \caption{Dynamical phase diagram of the post-quench Holstein model.
Shown is the spatiotemporal evolution of the local CDW order parameter $\phi_i(t)$ (top row) following an interaction quench at fixed adiabatic parameter $r=0.3$, for representative final electron–phonon couplings $\lambda$. The corresponding lattice displacements $Q_i(t)$ within the dashed regions are displayed in the bottom row. Three qualitatively distinct dynamical regimes are identified. Left: a nonequilibrium metallic regime ($\lambda < \lambda_{c1}$), in which CDW correlations remain short-ranged and fluctuate around zero, with no stable domain formation. Middle: a quasi-coarsening regime ($\lambda_{c1} < \lambda < \lambda_{c2}$), characterized by the nucleation of CDW domains separated by mobile kinks whose dynamics leads to slow, reaction-limited coarsening. Right: an arrested CDW regime ($\lambda > \lambda_{c2}$), where strong local CDW order develops rapidly, but kink motion is progressively suppressed at late times, resulting in frozen domain walls and dynamically arrested coarsening. From our numerical simulations, we estimate the critical couplings as $\lambda_{c1} \approx 0.4$ and $\lambda_{c2} \approx 1.0$.}
    \label{fig:dyn-phases}
\end{figure*}

In this section, we use the Ehrenfest non-adiabatic framework introduced in Sec.~II to investigate the nonequilibrium dynamics of the semiclassical one-dimensional Holstein model following an interaction quench. We focus on a deep-quench protocol in which the electron-phonon coupling is suddenly switched on from an initially decoupled state, $\lambda_i=0$, to a range of final values $\lambda_f$. The initial state is spatially uniform and symmetry unbroken: the electronic subsystem is a free spinless Fermi gas at half filling, while the lattice consists of independent harmonic oscillators prepared at rest, $Q_i=P_i=0$. In this limit, the two sectors are completely decoupled and no CDW order is present.   This homogeneous reference configuration provides a well-defined starting point for probing how strong electron--phonon coupling dynamically generates order and drives the system far from equilibrium. 

At time $t=0$, the electron-phonon coupling is abruptly quenched to a finite value $\lambda_f>0$. Because any nonzero coupling stabilizes a gapped CDW state in equilibrium, the pre-quench configuration constitutes a highly excited state of the post-quench Hamiltonian. The quench therefore injects a finite excess energy and initiates a strongly nonequilibrium evolution driven by the CDW instability. Following the quench, CDW order develops dynamically through a nucleation-and-growth process in which locally ordered regions emerge at random spatial locations and subsequently expand and coalesce into extended domains. This process bears a close resemblance to phase-ordering phenomena observed in thermal quenches, where an initially disordered system is driven into an ordered phase by suddenly coupling to a low-temperature reservoir. A crucial distinction, however, is that thermal quenches rely on energy dissipation to the environment, whereas in the present interaction quench the system is fully isolated and the total energy is strictly conserved, as discussed in Sec.~\ref{sec:holstein}.

To initiate symmetry-breaking dynamics in an otherwise perfectly homogeneous system, the lattice degrees of freedom are initialized with infinitesimal displacements and momenta of order $10^{-4}$. These perturbations do not represent external noise, stochastic forcing, or dissipation; rather, they act as controlled symmetry-breaking seeds that activate the intrinsic dynamical instability triggered by the sudden onset of the electron-phonon coupling. Their sole role is to select among symmetry-related ordering patterns, allowing the system to evolve deterministically under the post-quench Hamiltonian dynamics. To characterize the ensuing formation and evolution of CDW domains, we introduce a local CDW order parameter,
\begin{eqnarray}
\phi_i = \left( Q_i - \textstyle\sum_{j \in \mathcal{N}(i)} Q_j \right) (-1)^i ,
\end{eqnarray}
which measures the staggered component of the local density modulation, here $\mathcal{N}(i)$ denotes the two neighbors of site-$i$. This definition is tailored to detect the commensurate period-2 CDW order favored at half-filling and provides a convenient means to track domain growth and defect dynamics in real space. Within this controlled quench protocol, we identify three qualitatively distinct nonequilibrium dynamical regimes as a function of the final coupling strength $\lambda_f$: a nonequilibrium metallic regime without CDW order, an intermediate quasi-coarsening regime characterized by slow, scale-invariant domain growth, and a dynamically arrested CDW regime in which coarsening ceases and domain walls become immobile, as summarized in Fig.~\ref{fig:dyn-phases}.

\subsection{Nonequilibrium metal at weak coupling}

At weak electron-phonon coupling $\lambda < \lambda_{c1}$, the system does not exhibit CDW order, either locally or globally, at long times after the quench; see Fig.~\ref{fig:dyn-phases}(a). This absence of ordering is at first sight counterintuitive, since the zero-temperature ground state of the semiclassical Holstein model at half-filling is a commensurate period-2 CDW band insulator driven by Fermi-point nesting. A naive interpretation might therefore attribute the suppression of order to effective thermalization following the quench: the injected energy could, in principle, correspond to a high effective temperature at which CDW correlations---both long-ranged and local---are destroyed. However, several diagnostics rule out such a thermal scenario.

\begin{figure}
    \centering
    \includegraphics[width=0.99\columnwidth]{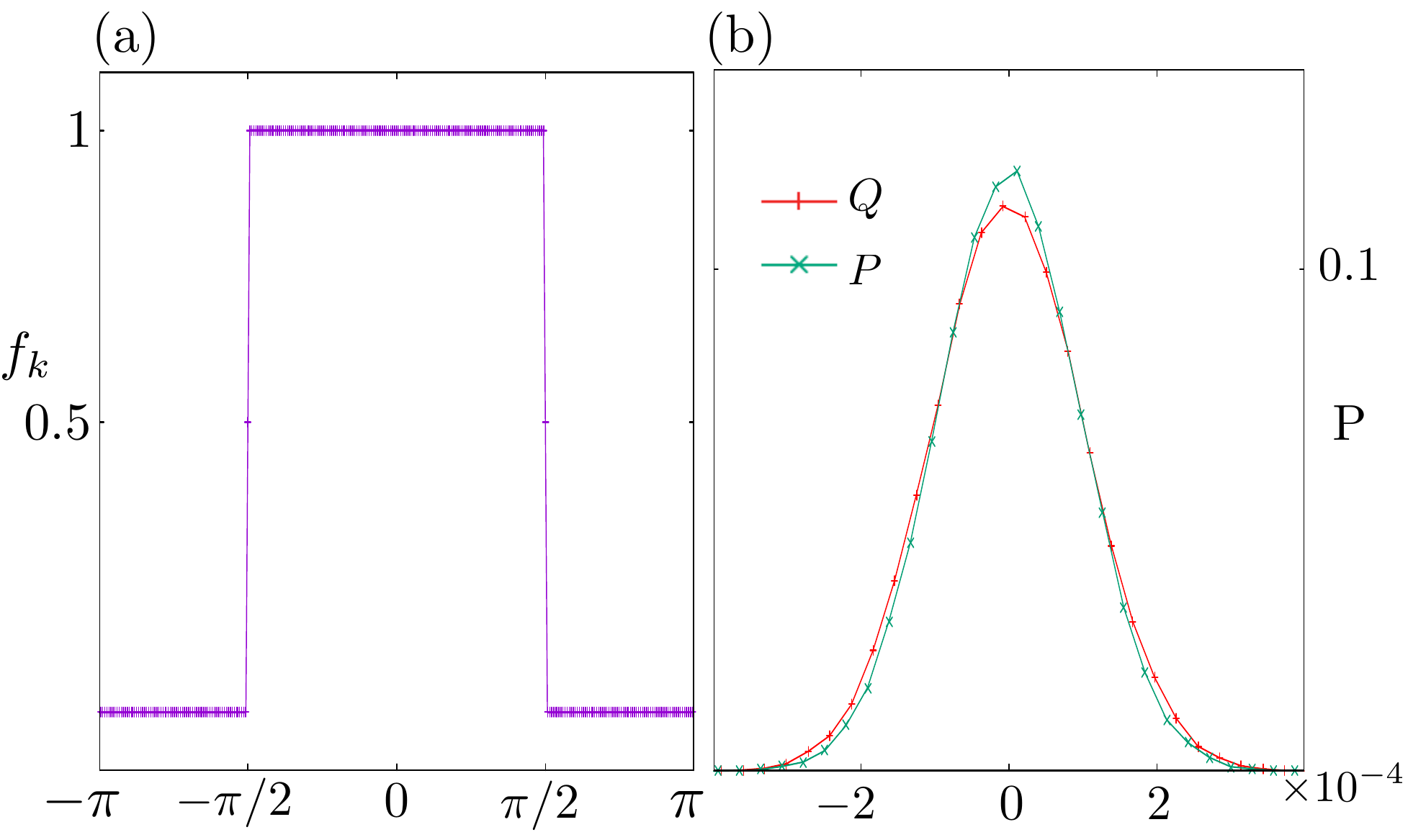}
    \caption{Nonequilibrium steady-state distributions in the weak-coupling metallic phase. (a) Late-time momentum-resolved electronic occupation $f_{\rm neq}(k)$, obtained from the single-particle density matrix in the nonadiabatic simulations. This distribution serves as the nonequilibrium analogue of the Fermi-Dirac function and retains a sharp Fermi surface, consistent with a metallic state. (b) Late-time distributions of the lattice displacement $Q$ and conjugate momentum $P$, which remain narrowly peaked and approximately Gaussian, indicating weak lattice fluctuations and the absence of CDW order.}
    \label{fig:neq_fk}
\end{figure}

To clarify the nature of this weak-coupling regime, we analyze the single-particle electronic density matrix $\rho(t)$ obtained from the nonadiabatic dynamics governed by Eq.~(\ref{eq:von_neumann_app}) by projecting onto the momentum eigenstates of the underlying tight-binding model. The resulting nonequilibrium momentum distribution is defined as
\begin{eqnarray}
	f_{\rm neq}(k, t) = {\rm Tr}\left[\rho(t) P_k \right].
\end{eqnarray}
where $P_k = |k\rangle\langle k|$ and $|k\rangle$ denotes a single-particle eigenstate of the tight-binding Hamiltonian. As shown in Fig.~\ref{fig:neq_fk}(a), the resulting late-time distribution $f_{\rm neq}(k)$ remains essentially indistinguishable from the initial near-zero-temperature Fermi-Dirac distribution, retaining a sharp discontinuity at the Fermi momentum and exhibiting no significant redistribution of spectral weight. Consistently, the distributions of lattice displacements $Q$ and conjugate momenta $P$ remain Gaussian with a narrow width of order $10^{-4}$, comparable to their initial values; see Fig.~\ref{fig:neq_fk}(b). Taken together, these observations demonstrate that the system does not experience appreciable heating and that, in this weak-coupling regime, the electronic and lattice subsystems remain effectively decoupled despite the sudden activation of the interaction.

\subsection{Arrested CDW states at strong coupling}

In the strong-quench regime, $\lambda > \lambda_{c2}$, the sudden activation of the electron-phonon coupling induces a rapid growth of local CDW order, as illustrated in Fig.~\ref{fig:dyn-phases}(c). The system quickly fragments into multiple CDW domains separated by kinks or domain walls. In sharp contrast to the weaker-coupling regimes, the ensuing evolution is qualitatively different: at late times the kinks become effectively immobile, and the domain pattern ceases to evolve. The system thus settles into a nonequilibrium steady state with frozen domain walls, which we refer to as an arrested CDW state. This behavior can be understood from the strong-coupling limit $g \gg t_{\rm nn}$, where electron hopping is strongly suppressed and the configuration energy approximately decomposes into a sum of independent on-site contributions, $E=\sum_i V(Q_i,n_i)$, with
\begin{eqnarray}
	V(Q_i,n_i)=-g(n_i-\tfrac{1}{2})Q_i+\tfrac{K}{2}Q_i^2. 
\end{eqnarray}
The resulting local potential has two nearly degenerate minima at $(Q=+Q^*, n=1)$ and $(Q=-Q^*, n=0)$, separated by a barrier of height $E_b \sim g^2/K$. Since each site can independently occupy either minimum, the system possesses an exponentially large manifold of nearly degenerate configurations. When finite hopping is restored, electronic delocalization weakly lifts this degeneracy and selects a CDW state with a doubled unit cell; however, the large barrier separating the local minima remains essentially unchanged.

Within this picture, kinks correspond to spatial regions where the local lattice distortion and charge configuration are trapped near one of the metastable minima of the now nondegenerate double-well potential. Crucially, the excess energy injected by the interaction quench---whether stored in electronic excitations or in lattice kinetic energy---is insufficient to overcome the barrier $E_b$ and activate kink motion or annihilation. As a consequence, defect dynamics is strongly suppressed, preventing further coarsening and locking the system into a long-lived, spatially inhomogeneous CDW configuration characterized by static domain walls and dynamically arrested evolution.

\subsection{Quasi-coarsening at intermediate coupling}

In the intermediate quench regime, $\lambda_{c1}<\lambda<\lambda_{c2}$, the post-quench dynamics displays pronounced coarsening behavior, as illustrated in Fig.~\ref{fig:dyn-phases}(b). Shortly after the quench, local CDW order develops and the system breaks up into multiple domains with opposite CDW phases, separated by kinks (domain walls). The subsequent evolution is governed by the motion and interaction of these defects, leading to a gradual increase of the characteristic domain size. Remarkably, this phenomenology closely resembles phase-ordering kinetics following a thermal quench, even though the present dynamics is strictly energy conserving and the system remains isolated throughout the evolution.

Despite this superficial similarity, the microscopic kink dynamics differs qualitatively from that in dissipative coarsening. Individual kinks primarily execute unbiased random-walk like motion, but their trajectories are intermittently punctuated by extended ballistic segments, reflecting the absence of friction and the persistence of inertia in the Hamiltonian dynamics. In addition, kink-kink collisions do not generically result in annihilation: instead of being irreversibly removed, kinks can scatter elastically and continue propagating after collision. Such elastic scattering processes, which are absent in overdamped thermal dynamics, significantly slow down the reduction of the defect density and place the coarsening in a reaction-limited regime. As a result, the intermediate phase is characterized by slow, non-diffusive domain growth that bridges the nonequilibrium metallic regime at weak coupling and the arrested CDW state at strong coupling.

\begin{figure}
    \centering
    \includegraphics[width=0.99\columnwidth]{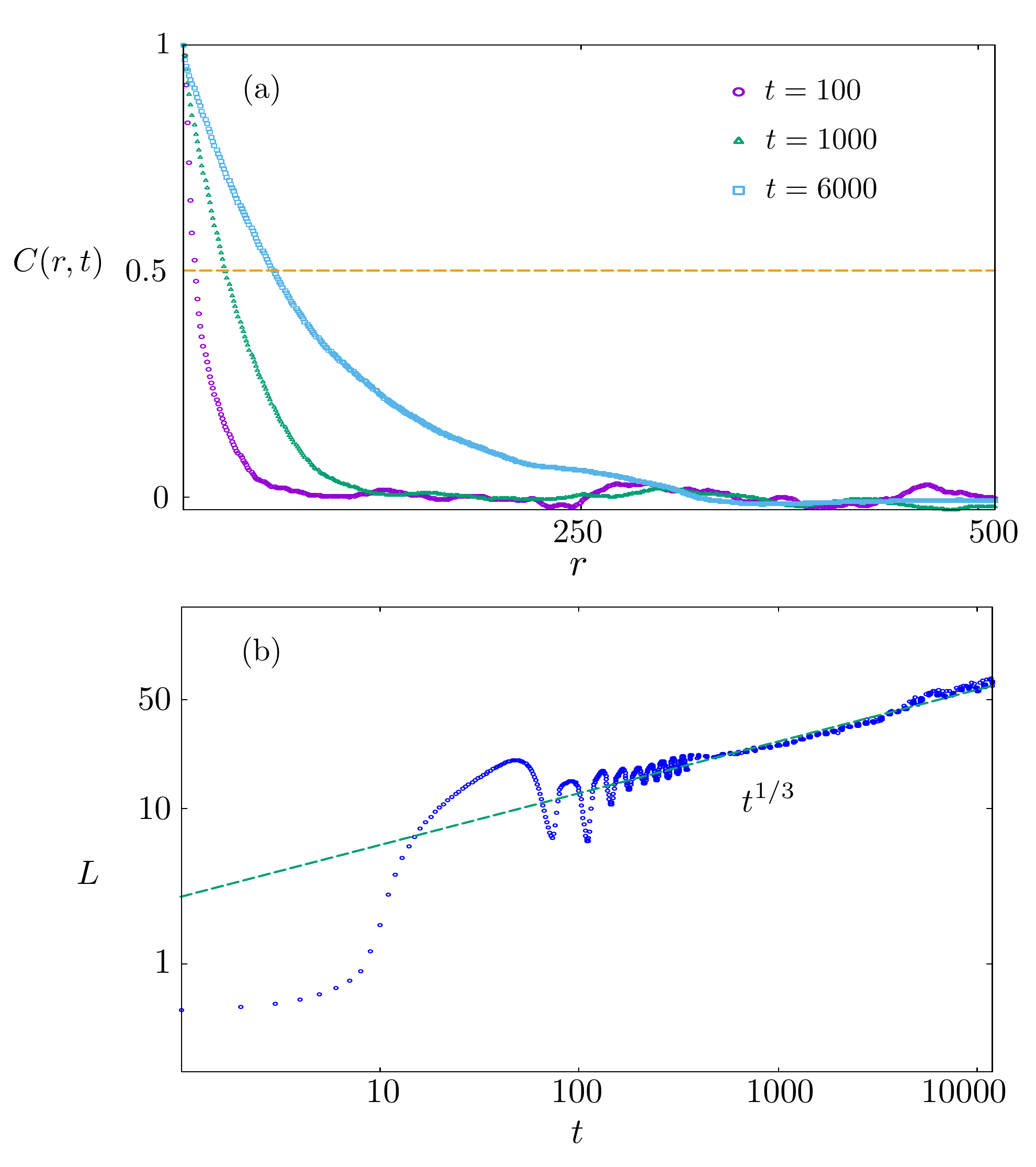}
    \caption{(a) Equal-time correlation function $C(r,t)$ of the local CDW order parameter, defined in Eq.~(\ref{eq:corr-func}), shown at representative times after the quench. Correlations progressively extend to larger length scales as time evolves. The dashed line indicates one half of the on-site value $C(0,t)$. (b) Time evolution of the correlation length $L(t)$, defined implicitly by $C(L,t)=\tfrac{1}{2}C(0,t)$. At late times, this characteristic length displays an algebraic growth $L(t)\sim t^{1/3}$, substantially slower than the diffusive coarsening expected for random-walk dynamics.
}
    \label{fig:correlation-function}
\end{figure}

To quantitatively characterize the growth of CDW domains following the interaction quench, we compute the equal-time correlation function of the staggered CDW order parameter $\phi_i$,
\begin{equation}
\label{eq:corr-func}
	C(r_{ij}, t) = \langle \phi_i(t)\phi_j(t) \rangle - \langle \phi_i(t) \rangle^2 ,
\end{equation}
where $r_{ij}=|i-j|$ denotes the lattice separation. Because the nucleation of CDW order is highly sensitive to microscopic fluctuations seeded at early times, all correlation functions are averaged over 100 independent realizations of the initial lattice displacements and momenta. The resulting correlation functions at representative times are shown in Fig.~\ref{fig:correlation-function}(a). At short distances, $C(r,t)$ decays rapidly, reflecting the formation of locally ordered CDW regions shortly after the quench. At larger separations, however, correlations remain weak and exhibit pronounced sample-to-sample fluctuations at early times, signaling the absence of long-range coherence between independently nucleated domains. As time evolves, correlations extend to progressively larger length scales, while the amplitude of these long-distance fluctuations is gradually suppressed, indicating the emergence of increasingly robust domain structure and a reduced sensitivity to the microscopic details of the initial conditions.

To extract a reliable measure of the characteristic domain size while minimizing the influence of irregular long-distance fluctuations, we introduce an empirical correlation length defined through a fixed-threshold criterion. Specifically, the correlation length $L(t)$ is simply identified as the distance at which the correlation function decays to half of its equal-site value, i.e. $C(L, t) = \frac{1}{2} C(0, t)$. This definition provides a stable and physically transparent characterization of domain growth across the entire temporal window. The resulting time dependence of $L(t)$ is shown in Fig.~\ref{fig:correlation-function}(b). Following an initial rapid increase associated with the onset of CDW ordering, $L(t)$ displays pronounced oscillations at intermediate times before crossing over to a well-defined asymptotic growth regime. At late times, the correlation length exhibits an algebraic scaling $L(t)\sim t^{1/3}$, indicating a slow coarsening process that is markedly subdiffusive. Similar oscillatory features have also been reported in experimental studies of CDW ordering, suggesting that they reflect intrinsic collective dynamics rather than numerical artifacts.

\begin{figure}
    \centering
    \includegraphics[width=0.9\columnwidth]{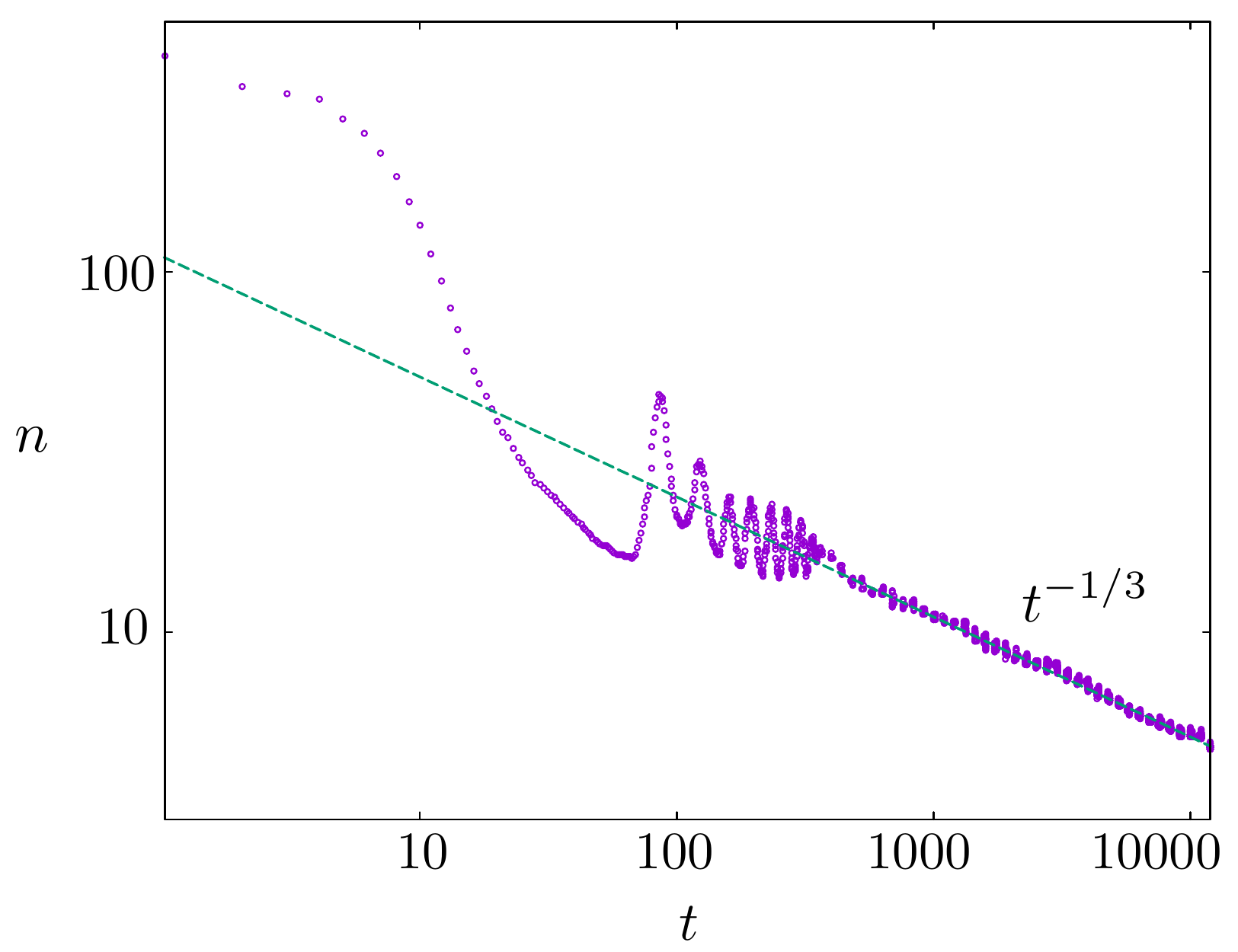}
    \caption{Kink decay in the quasi-coarsening regime.
Time evolution of the kink density $n(t)$ following the interaction quench. At late times, $n(t)$ exhibits an algebraic decay $n(t)\sim t^{-1/3}$ (dashed line). Over the full time window, this decay tracks the growth of the characteristic length scale via $L(t)\sim 1/n(t)$, indicating that the post-quench dynamics in the quasi-coarsening regime are governed by kink motion and interaction-driven annihilation.}
    \label{fig:n_t}
\end{figure}

At late times, when the correlation length becomes large compared to the microscopic lattice scale, a mean-field description in terms of topological defects becomes appropriate. In this regime, the characteristic length scale $L(t)$ can be interpreted as the typical separation between neighboring kinks (domain walls) separating CDW domains of opposite phase. As coarsening proceeds and domains merge, kinks become increasingly dilute, leading to a corresponding decay of the kink density $\rho(t)$. This picture naturally implies the scaling relation $L(t) \sim n^{-1}(t)$. This correspondence is directly demonstrated in Fig.~\ref{fig:n_t}, where the time evolution of the kink density is shown. In practice, kinks are identified using the standard CDW criterion: sites at which the local electronic density deviates above or below half filling relative to neighboring sites, signaling a phase slip of the staggered order parameter.

The close agreement between the growth of the correlation length and the decay of the kink density confirms that the post-quench dynamics in this regime are governed by the kinetics of kinks, including their motion, interactions, and annihilation events. The pronounced fluctuations observed at early and intermediate times reflect the strong sensitivity of domain nucleation to initial conditions and the complex interplay between kink propagation and scattering. At late times, these fluctuations are washed out, and the kink density approaches a clear asymptotic decay $n(t)\sim t^{-1/3}$, fully consistent with the scaling behavior extracted independently from the correlation function analysis. This consistency provides compelling evidence that the quasi-coarsening dynamics are controlled by defect-mediated processes in an energy-conserving, isolated system.

\begin{figure*}[t]
    \centering
    \includegraphics[width=0.96\linewidth]{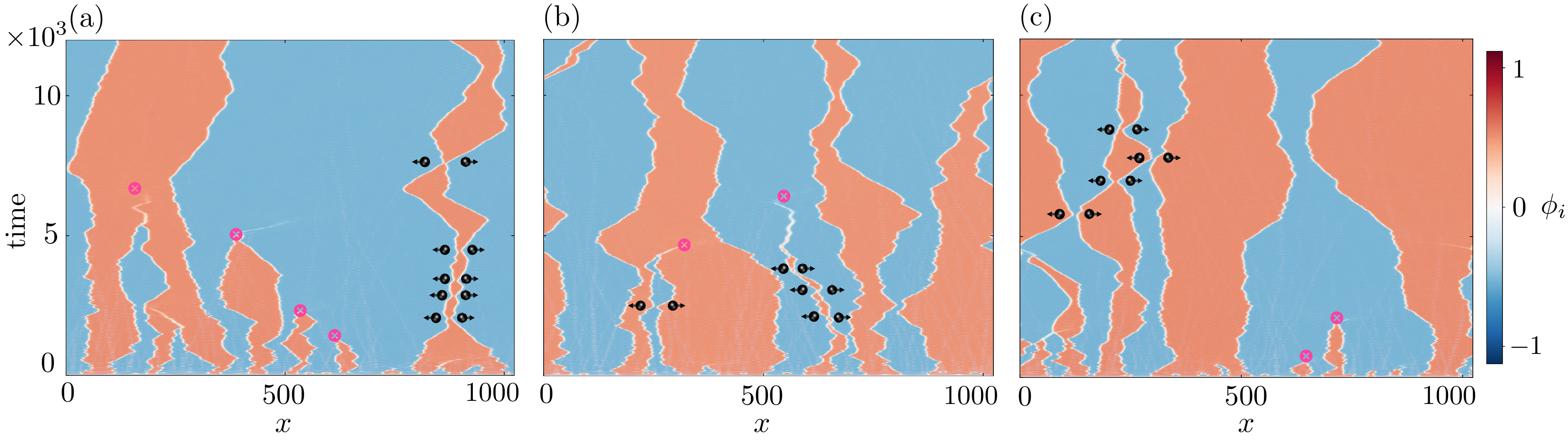}
    \caption{Space–time evolution of CDW domains and kink trajectories.
Representative space–time configurations of the local CDW order parameter in the quasi-coarsening regime, illustrating the dynamics of kinks (domain walls) separating CDW domains of opposite phase (indicated by contrasting colors). The kink trajectories are visible as the interfaces between domains. Magenta crosses mark kink-kink annihilation events, while pairs of black circles with outward-pointing arrows denote scattering events in which kinks collide but do not annihilate. The coexistence of annihilation and nonannihilating scattering highlights the reaction-limited nature of the coarsening dynamics.}
    \label{fig:snapshots}
\end{figure*}

\section{Effective Kinetics of Kinks in the Coarsening Regime}

In this section, we propose an effective kinetic description of kink dynamics to rationalize the anomalously slow coarsening exponent observed in our nonadiabatic simulations. In particular, we seek to explain the emergence of the algebraic growth law $L(t)\sim t^{1/3}$, or equivalently the decay of the kink density $n(t)\sim t^{-1/3}$, found in the quasi-coarsening regime. While the microscopic dynamics are governed by coupled quantum–classical equations of motion, the late-time evolution of the CDW order is naturally described in terms of a dilute gas of kinks separating locally ordered domains. The central goal of this section is to identify which effective kinetic processes control the long-time evolution of these defects and how they differ from conventional coarsening scenarios.

A natural starting point is diffusion-limited annihilation of kinks, which underlies the standard coarsening behavior of nonconserved Ising order parameters. In this picture, kinks perform unbiased random walks due to thermal or stochastic fluctuations and annihilate upon encounter. The typical separation between neighboring kinks grows diffusively, $L(t)\sim t^{1/2}$, reflecting the scaling of the root-mean-square displacement of a random walker. Correspondingly, the kink density decays as $n(t)\sim t^{-1/2}$. This mechanism provides a robust and well-established description of phase ordering in dissipative systems with purely relaxational dynamics. However, the clear deviation of our results from the $1/2$ exponent rules out simple diffusion-limited annihilation as the dominant late-time process.

A $1/3$ coarsening exponent is instead commonly associated with Kawasaki dynamics, which describes phase ordering of Ising systems with a conserved order parameter. In that case, domain growth is controlled by diffusive transport of the conserved quantity across domain walls, leading to $L(t)\sim t^{1/3}$. Importantly, this mechanism relies on local conservation laws that constrain defect motion and slow down coarsening relative to the nonconserved case. While the numerical exponent observed in our simulations coincides with this classical result, the underlying physical setting is fundamentally different. In particular, the staggered CDW order parameter in our system is not conserved, nor is the dynamics governed by diffusive exchange processes characteristic of Kawasaki kinetics. The appearance of a $1/3$ exponent in our simulations therefore cannot be attributed to conserved-order-parameter dynamics.

Previous studies of Hamiltonian quenches in isolated systems have shown that kink motion can deviate strongly from diffusive behavior and instead acquire a ballistic character, either due to inertia or as a consequence of strict energy conservation. In such settings, kinks propagate over long distances with well-defined velocities, and their dynamics are only weakly randomized by interactions with other excitations. These observations naturally place the coarsening dynamics outside the standard diffusion-limited framework and invite comparison with earlier studies of annihilation processes involving ballistic or biased defect motion in one dimension.

Well before the context of Hamiltonian quenches, annihilation dynamics of ballistically moving particles were studied extensively as an independent class of reaction–diffusion models~\cite{BenNaim1996,Biswas2016,Biswas2021}. In the symmetric case of equal densities of left- and right-moving particles with fixed velocities, annihilating upon contact, particle motion combines deterministic ballistic propagation with an effective stochastic component arising from asynchronous updates or diffusive noise. Ballistic motion rapidly brings oppositely moving particles into contact, while the stochastic component enables annihilation between particles moving in the same direction by allowing their relative separation to fluctuate. This combination leads to self-organization into domains of uniform velocity separated by sharp interfaces, with annihilation occurring both at domain boundaries and within domains. As a result, the particle density decays as $n(t)\sim t^{-3/4}$~\cite{Biswas2021}, substantially faster than the $t^{-1/2}$ law characteristic of purely diffusive annihilation.

More generally, these studies establish that ballistic transport generically enhances annihilation kinetics and leads to coarsening exponents exceeding $1/2$. This trend is consistent with recent simulations of isolated Hamiltonian quenches in 1D $\phi^4$ chains~\cite{bastianello2020}, where kinks propagate ballistically and annihilation is sufficiently efficient to yield an even faster decay, $n(t)\sim t^{-1}$. Together, these results underscore that ballistic kink motion alone tends to accelerate coarsening, reinforcing the conclusion that the slow $t^{-1/3}$ scaling observed in our system must arise from a different physical mechanism, namely the suppression of annihilation due to elastic scattering in an otherwise diffusive kink gas.

Representative space-time configurations of the CDW order parameter in the quasi-coarsening regime are shown in Fig.~\ref{fig:snapshots}. These snapshots reveal that, at intermediate and late times after the interaction quench, kink trajectories are predominantly unbiased and diffusive, with no persistent directional drift. While short-time segments of individual trajectories can appear weakly ballistic, such behavior does not persist at long times. This crossover can be understood by viewing the electronic degrees of freedom as an effective dynamical reservoir: incoherent scattering between kinks and the electronic background continually randomizes kink velocities, leading to an overall unbiased random-walk motion despite the absence of external dissipation. Importantly, when two kinks encounter each other, two distinct outcomes are observed. In some cases, the kinks annihilate, as marked by magenta crosses in the figure, while in others they undergo elastic scattering and separate without annihilation, indicated by pairs of black circles with outward-pointing arrows. The coexistence of these two processes is a central feature of the quasi-coarsening regime.

To capture these observations at a coarse-grained level, we introduce an effective kinetic model for kink motion. In this model, kinks are treated as point-like particles undergoing unbiased random walks, representing their long-time diffusive dynamics mediated by interactions with the electronic background. When two kinks encounter each other, the collision is stochastic and admits two possible outcomes,
\begin{eqnarray} 
A + A \rightarrow \left\{\begin{array}{ll} \emptyset & \qquad \mbox{with probability $p$} \\ A + A & \qquad \mbox{with probability $1-p$} \end{array} \right. 
\end{eqnarray}
corresponding respectively to annihilation and elastic scattering. To model the post-collision separation following a scattering event, we assume that the two kinks temporarily execute biased random walks for a finite duration $n_{\rm bias}$, with opposite biases that drive them away from each other. The bias is controlled by a parameter $b$, such that the probability to move in the outward direction is $1/2+b$, while motion in the opposite direction occurs with probability $1/2-b$. After this transient bias window, the kinks revert to unbiased diffusion. This minimal model incorporates both suppressed annihilation and elastic scattering, providing a natural framework for understanding the reaction-limited coarsening dynamics observed in the simulations.

\begin{figure}
    \centering
    \includegraphics[width=0.96\columnwidth]{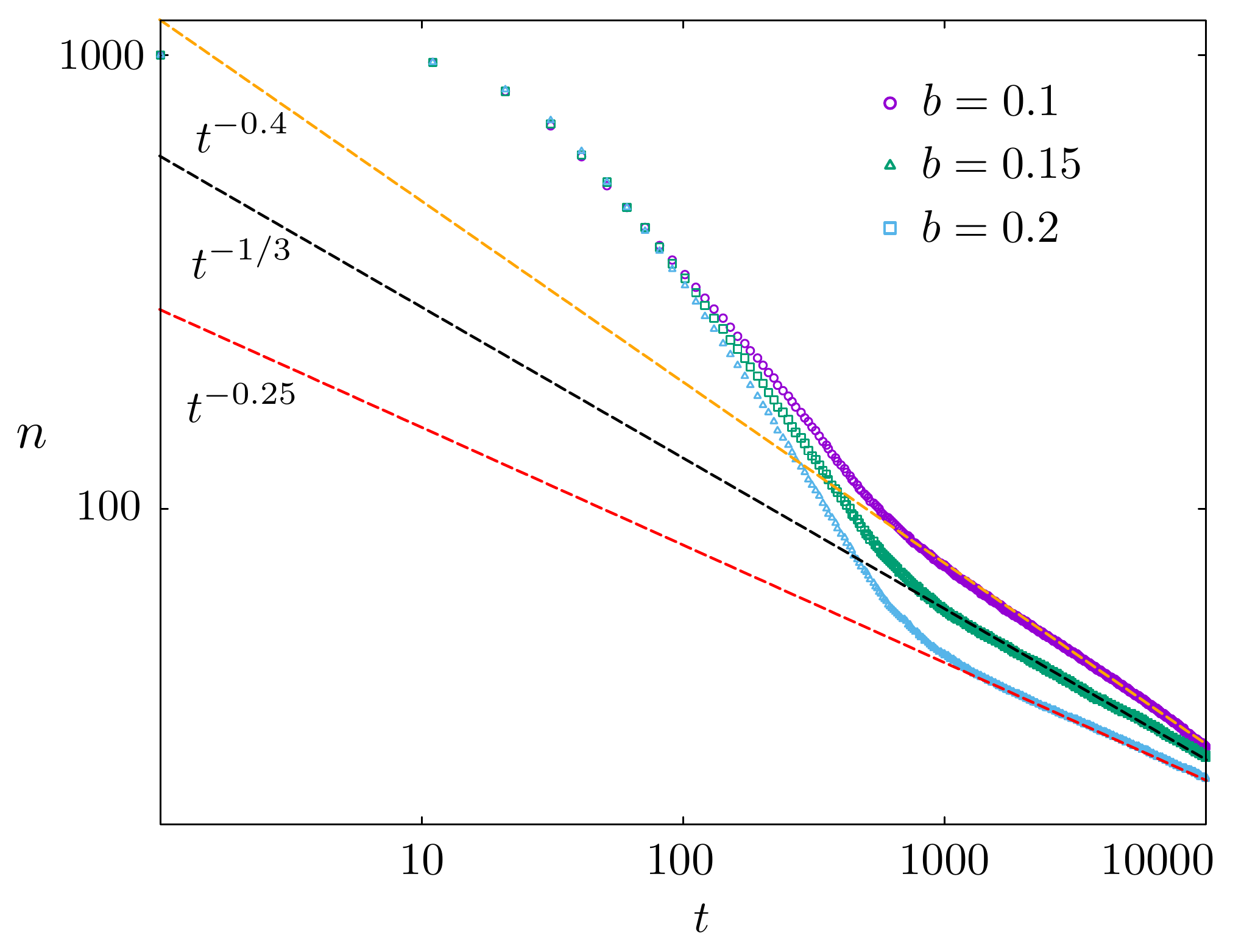}
    \caption{Large-scale simulations of the effective kinetic kink model.
Time evolution of the kink density $n(t)$ for different values of the scattering parameter $b$, which controls the relative probability of elastic kink–kink scattering versus annihilation. Increasing $b$ leads to a faster decay at intermediate times but a slower asymptotic decay at late times. The resulting scaling behavior spans a range of exponents that includes the $t^{-1/3}$ decay observed in the quasi-coarsening regime of the nonadiabatic quench dynamics.}
    \label{fig:kink-kinetics}
\end{figure}

We simulate the effective kinetic model by initializing $N=1000$ kinks at random positions on a one-dimensional ring of length $L=10^4$, and follow their stochastic evolution under unbiased diffusion interrupted by collision-induced annihilation or scattering. The kink density $n(t)$ is obtained by averaging over 300 independent realizations, where $t$ denotes the number of update iterations. The resulting time evolution of $n(t)$ is shown in Fig.~\ref{fig:kink-kinetics} for several values of the scattering bias parameter $b$. Across all cases, the dynamics exhibit a pronounced two-stage structure, reflecting a crossover between dense and dilute defect regimes.

At early and intermediate times, when the kink density is high and collisions are frequent, increasing the bias strength $b$ leads to a faster initial decay of $n(t)$. In this regime, post-collision biased motion enhances the spatial separation of kinks after scattering events, reducing repeated near encounters and effectively increasing the rate at which distinct kink pairs are explored. As a result, annihilation events are more efficiently sampled, producing a transient coarsening behavior with a larger effective decay exponent. This bias-enhanced early-time decay is evident in Fig.~\ref{fig:kink-kinetics}, where stronger bias produces steeper slopes at intermediate times.

At late times, however, once the system enters a dilute regime characterized by large inter-kink separations, the effect of bias is reversed. In this regime, scattering events dominate over annihilation, and the post-collision biased motion systematically drives kinks away from one another, thereby suppressing subsequent encounters. Consequently, annihilation becomes reaction-limited rather than transport-limited, and the decay of $n(t)$ slows down markedly. As shown in Fig.~\ref{fig:kink-kinetics}, increasing $b$ reduces the magnitude of the asymptotic decay exponent, producing progressively slower coarsening at long times. Notably, for intermediate bias values, the late-time decay crosses over to an effective exponent close to $1/3$, consistent with the scaling observed in the quasi-coarsening regime of the full nonadiabatic quench dynamics. These results demonstrate that the competition between unbiased diffusion, elastic scattering, and suppressed annihilation naturally generates slow coarsening without invoking conserved order parameters or purely ballistic transport.

To further isolate the mechanism responsible for the slow coarsening, we also examined an alternative effective kinetic model in which kinks undergo biased random walks, mimicking persistent or directed (ballistic) motion. In this variant, each kink is assigned an internal label $\sigma=\pm 1$ that specifies its preferred direction of motion. Upon encounter, two kinks annihilate with probability $p$, or scatter elastically with probability $1-p$, in which case their direction labels $\sigma$ are reversed. Despite incorporating the same stochastic annihilation–scattering structure, extensive simulations of this biased-walk model consistently yield decay exponents exceeding $1/2$, indicating faster coarsening than observed in the Holstein nonadiabatic dynamics. This outcome underscores a key distinction: persistent directional motion enhances encounter rates even when annihilation is probabilistic, and therefore cannot account for the reaction-limited kinetics underlying the $t^{-1/3}$ scaling. These results highlight that unbiased diffusive motion of kinks---arising from incoherent interactions with the electronic background---is essential for suppressing annihilation and realizing the slow quasi-coarsening regime observed in the full microscopic simulations.

\section{Conclusion and outlook}

To summarize, we have investigated the nonequilibrium dynamics induced by an interaction quench in the one-dimensional semiclassical Holstein model within the Ehrenfest nonadiabatic framework, which describes an isolated hybrid quantum-classical system with strictly conserved total energy. Through large-scale real-time simulations, we identified three qualitatively distinct dynamical regimes as a function of the post-quench electron–phonon coupling: a nonequilibrium metallic phase, an intermediate quasi-coarsening regime, and a strongly coupled arrested CDW state. In the quasi-coarsening regime, CDW order emerges through the nucleation and slow growth of domains separated by mobile kinks, yielding dynamics that closely resemble phase ordering despite the absence of dissipation. Analysis of correlation functions and defect statistics revealed an unconventional algebraic scaling characterized by a correlation length $L(t)\sim t^{1/3}$ and a kink density decay $n(t)\sim t^{-1/3}$, distinct from both diffusive coarsening in dissipative systems and ballistic annihilation in isolated Hamiltonian chains. These results point to a novel coarsening mechanism in which electronic degrees of freedom act as an internal reservoir, inducing diffusive kink motion without energy loss.

To provide a minimal theoretical interpretation of this behavior, we developed an effective kinetic description of kink dynamics that captures the essential late-time processes observed in the microscopic simulations. In this framework, kinks undergo unbiased random walks and interact through stochastic annihilation and elastic scattering upon encounters. Large-scale simulations of the effective kinetics reveal a two-stage evolution, with an early regime of enhanced annihilation followed by a reaction-limited late-time regime in which elastic scattering suppresses further annihilation and produces slow algebraic decay. Crucially, we showed that unbiased diffusive motion is essential for realizing the $1/3$ scaling: models incorporating persistent directional bias or ballistic motion invariably lead to faster coarsening, inconsistent with the Holstein dynamics. The close quantitative agreement between the effective kinetic model and the full nonadiabatic simulations supports a unified picture of slow quasi-coarsening in isolated hybrid systems, and suggests broader relevance for reaction-limited defect kinetics in nonequilibrium quantum materials.

An important open direction concerns the microscopic origin of the effective kinetic parameters introduced in this work. In particular, a first-principles understanding of the annihilation probability $p$, the post-collision bias parameter $b$, and the associated timescale $n_{\rm bias}$ remains to be developed. These quantities encode how energy and momentum are redistributed between kinks and electronic degrees of freedom during close encounters, and are therefore expected to depend sensitively on the local electronic spectrum, phonon stiffness, and quench protocol. A systematic derivation of these parameters from the underlying nonadiabatic dynamics---potentially via controlled few-kink simulations, scattering theory in time-dependent backgrounds, or reduced descriptions based on electronic response functions---would provide a direct microscopic bridge between the Holstein model and the effective kinetic framework. Such an approach would also clarify the extent to which the effective kinetics is universal or model-dependent.

Beyond the intermediate quasi-coarsening regime emphasized here, a central open issue concerns thermalization in the post-quench dynamics of the Holstein model. The Ehrenfest framework employed in this work is tailored to describe coherent evolution of quadratic fermions coupled self-consistently to classical phonons, and therefore lacks generic mechanisms for ergodic energy redistribution. As a result, the system is not expected to thermalize in the conventional sense, a fact already borne out by the persistence of the nonequilibrium metallic state following weak interaction quenches. Instead, the post-quench dynamics explored here is naturally interpreted in terms of prethermalization, where the evolution becomes trapped for long times within a restricted region of phase space due to emergent dynamical constraints. Within this perspective, the three dynamical regimes identified in this study likely correspond to stable or long-lived prethermal states of the isolated system. A key direction for future work is to examine how these regimes are modified or destabilized when genuine thermalization channels are introduced, for instance by incorporating quantum fluctuations of the phonons beyond the semiclassical approximation or by including electron-electron interactions. Such extensions will be crucial for determining which aspects of the observed defect kinetics and coarsening behavior survive in more realistic interacting settings, and for clarifying the interplay between prethermal dynamics and eventual thermal equilibration.

\begin{acknowledgments}
This work was supported by the Owens Family Foundation. The authors thank Lingyu Yang for assistance during the early stages of this project. Computational resources and technical support were provided by Research Computing at the University of Virginia.
\end{acknowledgments}

\appendix

\section{Dimensional analysis}

The semiclassical Ehrenfest equations derived above contain multiple intrinsic energy and time scales associated with electronic motion, lattice dynamics, and electron--phonon coupling. To elucidate their interplay and to identify the minimal set of control parameters governing the nonadiabatic regime, it is useful to express the equations of motion in dimensionless form.

We begin by fixing the electronic kinetic energy scale through the nearest-neighbor hopping amplitude $t_{\rm nn}$, which defines a natural electronic timescale
\begin{equation}
    \tau_e =\frac{\hbar}{t_{\rm nn}} .
\end{equation}
Time is then expressed in electronic units by introducing the dimensionless variable $\tilde{t}= t/\tau_e$; correspondingly, all energies are reported in units of $t_{\rm nn}$.


A characteristic amplitude for lattice distortions follows from the competition between the elastic restoring force and the local electron--phonon coupling. Equating the elastic energy $K Q^2$ with the coupling energy $g Q$ yields the displacement scale
\begin{equation}
    Q^* = \frac{g}{K}.
\end{equation}
We therefore rescale lattice displacements according to $\tilde{Q}_i = Q_i/Q^*$. In terms of $\tilde{t}$ and $\tilde{Q}_i$, the electronic sector evolves under a dimensionless single-particle Hamiltonian whose matrix elements depend only on rescaled lattice fields.

The lattice dynamics introduces an independent timescale through the bare phonon frequency $\Omega$, or equivalently
\begin{equation}
    \tau_L = \frac{1}{\Omega} = \sqrt{\frac{m}{K}} .
\end{equation}
The ratio between electronic and lattice timescales,
\begin{equation}
    r = \frac{\tau_e}{\tau_L} = \frac{\hbar \Omega}{t_{\rm nn}},
\end{equation}
quantifies the degree of nonadiabaticity. In the limit $r \ll 1$, lattice motion is slow compared to electronic dynamics and the system approaches the adiabatic regime, whereas $r \sim \mathcal{O}(1)$ corresponds to strongly nonadiabatic dynamics with comparable electronic and lattice timescales.

To write the lattice dynamics in a form that cleanly exposes the two intrinsic timescales, it is convenient to introduce a dimensionless lattice momentum. A natural momentum scale follows from harmonic motion, $\dot{Q}\sim \Omega Q$, which suggests
\begin{equation}
    P^* = m \Omega Q^* .
\end{equation}
With $\tilde{P}_i \equiv P_i/P^*$, the lattice equations of motion become
\begin{eqnarray}
\label{dimless_latticeEOM}
    \frac{d\tilde{Q}_i}{d\tilde{t}} &=& r\, \tilde{P}_i, \nonumber \\
    \frac{d\tilde{P}_i}{d\tilde{t}} &=& - r\, \tilde{Q}_i
    + r\left( n_i - \frac{1}{2} \right),
\end{eqnarray}
where $n_i$ is the electronic occupation at site $i$. In the nonadiabatic regime, the electronic sector is evolved explicitly alongside the lattice, via the dimensionless von Neumann equation.

The dimensionless von Neumann equation for the single-particle density matrix becomes
\begin{equation}
    i \frac{d\rho_{ij}}{d\tilde{t}}
    =
    \sum_k \left( \rho_{ik} \tilde{t}_{kj} - \tilde{t}_{ik} \rho_{kj} \right)
    + 4\lambda \left( \tilde{Q}_j - \tilde{Q}_i \right)\rho_{ij},
    \label{eq:A2}
\end{equation}
where $\tilde{t}_{ij} = t_{ij}/t_{\rm nn}$. The dimensionless coupling constant
\begin{equation}
    \lambda = \frac{g^2}{K W},
\end{equation}
defined in terms of the electronic bandwidth $W = 4 t_{\rm nn}$, parametrizes the strength of electronic backaction on the lattice and controls the stability of charge-density-wave order at the semiclassical level. The electron occupation $n_i(=\rho_{ii})$ evolved from the dimensionless von Neumann equation couples into the lattice dynamics.

Taken together, Eqs.~(\ref{dimless_latticeEOM}) and (\ref{eq:A2}) define a closed, dimensionless dynamical system controlled by two independent parameters: the nonadiabaticity parameter $r$ and the electron--phonon coupling strength $\lambda$. These parameters span the phase space of semiclassical Holstein dynamics and determine both the stability of ordered states and the character of electron--lattice feedback far from equilibrium. The dynamical evolution has been performed numerical integration with fourth order Runge-Kutta method.

\bibliography{ref}

\end{document}